\documentclass[prb,twocolumn,showpacs,preprintnumbers,amsmath,amssymb,superscriptaddress, bibnotes]{revtex4}

\usepackage{color}
\usepackage{graphicx}
\usepackage{dcolumn}
\usepackage{bm}
\usepackage{mathrsfs}
\usepackage{tabls}
\newcommand{\BFAP}{BaFe$_2$(As$_{1-x}$P$_{x}$)$_2$}
\newcommand{\BFCA}{Ba(Fe$_{1-x}$Co$_{x}$)$_2$As$_2$}

\begin{document}

\title{Normal state spin dynamics in the iron-pnictide superconductors \\
\BFAP \ and \BFCA \ probed with NMR measurements}

\author{Y.~Nakai}
\email{nakai@tmu.ac.jp}
\affiliation{Department of Physics, Graduate School of Science and Engineering, Tokyo Metropolitan University, Tokyo 192-0397, Japan,}
\author{T.~Iye}
\author{S.~Kitagawa}
\author{K.~Ishida}
\email{kishida@scphys.kyoto-u.ac.jp}
\affiliation{Department of Physics, Graduate School of Science, Kyoto University, Kyoto 606-8502, Japan,}
\affiliation{TRIP, JST, Sanban-cho, Chiyoda, Tokyo 102-0075, Japan,}

\author{\\S.~Kasahara}
\affiliation{Department of Physics, Graduate School of Science, Kyoto University, Kyoto 606-8502, Japan,}
\affiliation{Research Center for Low Temperature and Materials Sciences, Kyoto University, Kyoto 606-8502, Japan}
\author{T.~Shibauchi}
\author{Y.~Matsuda}
\author{H.~Ikeda}
\affiliation{Department of Physics, Graduate School of Science, Kyoto University, Kyoto 606-8502, Japan,}
\author{T.~Terashima}
\affiliation{Research Center for Low Temperature and Materials Sciences, Kyoto University, Kyoto 606-8502, Japan}

\date{\today}

\begin{abstract}
The NMR results in iron pnictides \BFAP \  and \BFCA \ are analyzed based on the self-consistent renormalization (SCR) spin fluctuation theory. 
The temperature dependence of the NMR relaxation rate $T_1^{-1}$ as well as the electrical resistivity is well reproduced by a SCR model where two-dimensional antiferromagnetic (AF) spin fluctuations are dominant. 
The successful description of the crossover feature from non-Fermi liquid to Fermi liquid behavior strongly suggests that low-lying spin fluctuations in \BFAP \  and \BFCA \ possess an itinerant AF nature, 
and that chemical substitution in the two compounds tunes the distance of these systems to an AF quantum critical point. 
The close relationship between spin fluctuations and superconductivity is discussed compared with the other unconventional superconductors, cuprate and heavy fermion superconductors. In addition, it is suggested that magnetism and lattice instability in these pnictides are strongly linked via orbital degrees of freedom. 
\end{abstract}

\pacs{76.60.-k,	
74.25.nj, 
74.40.Kb, 
74.70.Xa 
}
\maketitle

\section{Introduction}
Since the discovery of high-temperature superconductivity in iron-pnictide superconductors, much effort has been paid to the understanding of the normal and superconducting (SC) state properties, and considerable interest has been focused on the origin of the pairing interaction.~\cite{KamiharaFeAs,IshidaJPSJReview,PaglioneNature} 
The proximity of a SC to an antiferromagnetic (AF) phase strongly suggests the interplay between the two ground states. 
There is accumulating evidence that AF quantum criticality is deeply related to the physics of iron-pnictide superconductors.~\cite{JiangBaFe2(AsP)2,KasaharaBaFe2(AsP)2,NingPRL2010,NakaiPRL2010,IyeJPSJ2012,IyePRB2012,JDaiPNAS2009,SachdevPhysicsToday2011} 
We have studied the spin fluctuations in \BFAP \ with NMR measurements, and showed that AF spin fluctuations strongly correlate with superconductivity in this system,~\cite{NakaiBaFe2(AsP)2,NakaiPRL2010} where a line-nodal SC gap structure is suggested.~\cite{HashimotoPenetrationBaFe2(AsP)2,NakaiBaFe2(AsP)2,KimSpecificHeatBaFe2(AsP)2,YamashitaPRB2011} 
These NMR measurements, as well as dHvA experiments, suggest that AF spin fluctuations with a quantum critical nature could be responsible for the ``glue" that binds the SC Cooper pairs. 
In addition, ``quantum critical'' behavior was reported in a SC-parameter, London penetration  depth $\lambda_{\rm L}$, which is direct experimental evidence that the superconductivity in \BFAP \ is linked with its magnetic properties.~\cite{HashimotoScience2012} 

In this paper, we analyze in more detail experimental results, particularly the NMR relaxation rate of \BFAP \ and  \BFCA. 
They both possess the ``122'' structure (ThCr$_2$Si$_2$ structure) and the recent thermal expansion experiment showed their thermodynamic similarity.~\cite{BohmerPRB2012} 
These compounds are suggested to be close to an AF quantum critical point (QCP),~\cite{NakaiPRL2010,NingPRL2010} on the basis of the self-consistent renormalization (SCR) theory of  spin fluctuations. 
The SCR theory, developed by Moriya and coworkers, has been applied for weak ferromagnetism and antiferromagnetism of $d$-electron itinerant magnets, and succeeded in characterizing properties of spin fluctuations. 
As recent studies have shown the importance of both the itinerant and localized nature of the magnetism of iron pnictides,~\cite{DaiNaturePhys2012,GretarssonPRB2011,VilmercatiPRB2012,GorkovPRB2013} it is important to show to what extent experimental results are understood within an itinerant and local-moment picture. 
Recently, x-ray emission spectroscopy, which is sensitive to very rapid time scales, allowed for the detection of large local moments in the paramagnetic states in iron pnictides.~\cite{GretarssonPRB2011,VilmercatiPRB2012}
In contrast, NMR is a very useful probe to detect much slower fluctuations or low-energy spin excitation, enabling us to extract the itinerant aspects of iron pnictides. 

We derive spin-fluctuation parameters in the two compounds by taking into account other experimental results such as the static magnetic susceptibility and specific heat. 
We calculate the temperature dependencies of the NMR relaxation rate and the electrical resistivity following the SCR theory, and show that our calculations are quantitatively consistent with the experimental data. 
Our analysis indicates that the $T_c$ maximum concentration corresponds to an AF QCP and suggests the possibility of magnetically mediated high-$T_c$ superconductivity in the ``122'' iron-pnictide superconductors as in other unconventional superconductors of strongly-correlated-electron systems.

\section{Survey of NMR experiments}
Most of our NMR experimental results in BaFe$_2$(As$_{1-x}$P$_x$)$_2$ were published.~\cite{NakaiPRL2010, IyeJPSJ2012, IyePRB2012} 
In order to reanalyze our published NMR data in terms of the SCR theory, we summarize them in Figs.~1 and 2.

Figure~1 demonstrates the temperature and P-concentration dependence of the Knight shift in \BFAP, which is a measure of the static spin susceptibility $\chi({\bm q} = 0)$. 
The Knight shift is basically $T$-independent, but P substitution reduces the magnitude of the Knight shift.@
These results are attributable to the decrease in the density of states (DOS) at the Fermi level with P substitution.~\cite{IkedaJPSJ2008} 
\begin{figure}[tb]
\begin{center}
\includegraphics[width=8cm]{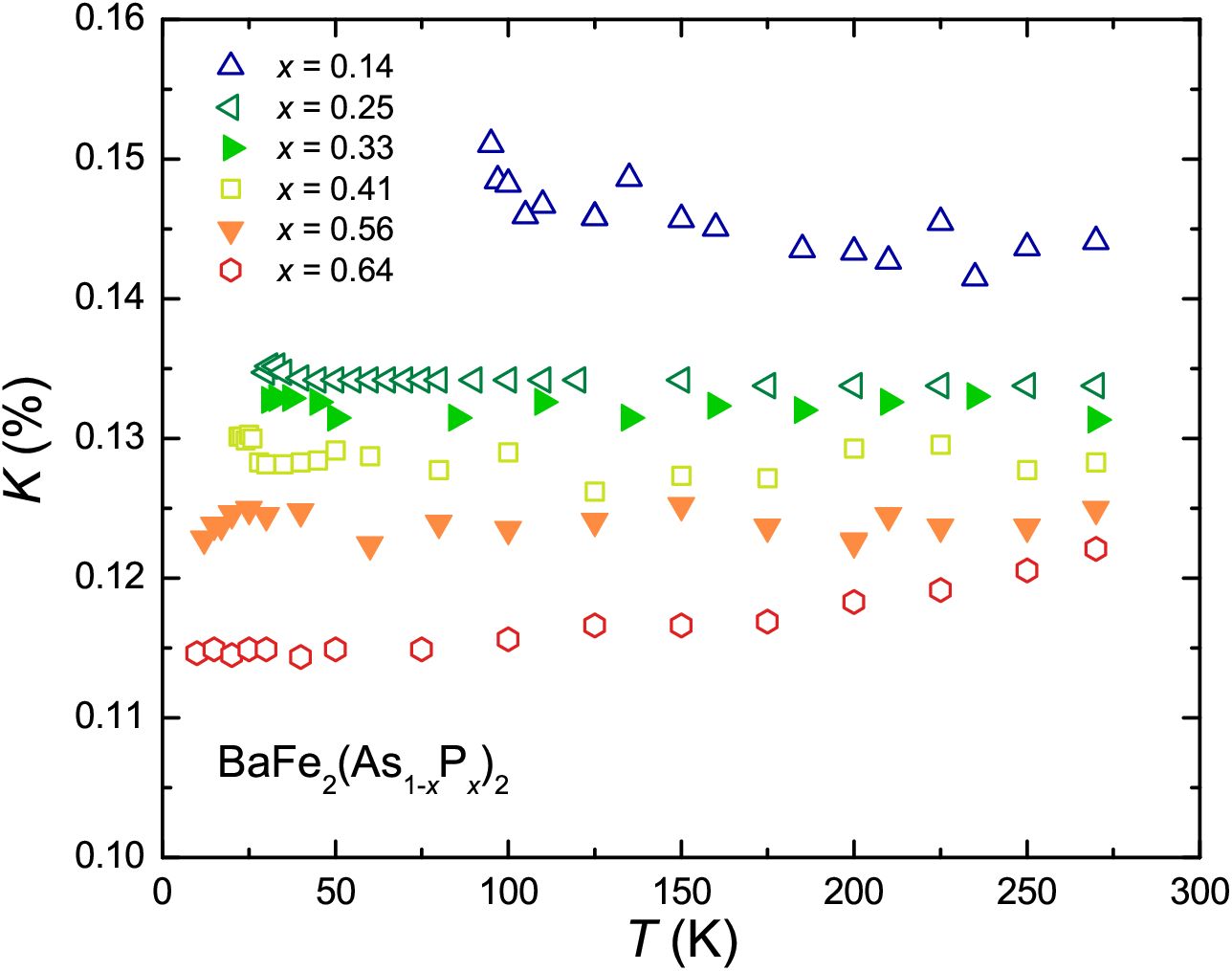}
\end{center}
\caption{(Color online) The Knight shift of the $^{31}$P nucleus for different P concentration of \BFAP.}
\label{}
\end{figure}

\begin{figure}[tb]
\begin{center}
\includegraphics[width=8.5cm]{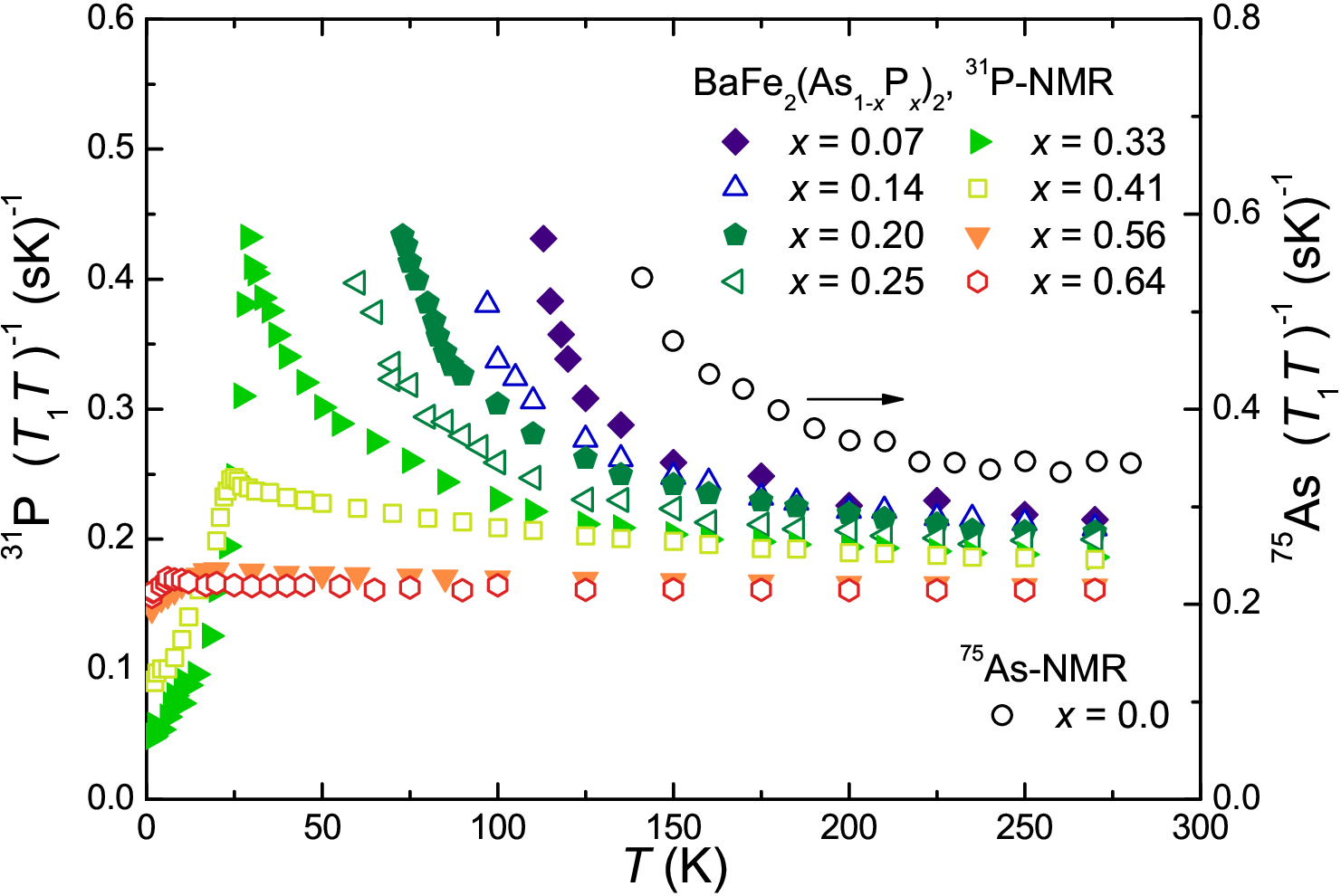}
\end{center}
\caption{(Color online) The $^{31}$P nuclear spin-lattice relaxation rate divided by temperature $(T_1T)^{-1}$ for \BFAP. The data for $x=0$ (BaFe$_2$As$_2$) were cited from Ref.~21.}
\label{T1T}
\end{figure}

Figure~\ref{T1T} displays the temperature dependence of $(T_1T)^{-1}$ for \BFAP \ with various P-concentration, where P substitution suppresses antiferromagnetism and induces superconductivity. 
We observe non-Fermi-liquid (NFL) temperature dependence of the Curie-Weiss (CW) form $(T_1T)^{-1} = a + b/(T+\theta)$ in the paramagnetic temperature range.  
For $x\le 0.20$, $(T_1T)^{-1}$ increases on cooling and has a peak at $T_{\rm N}$ due to the opening of spin density wave gap, but for $x \ge 0.33$, $(T_1T)^{-1}$ exhibits a peak due to a SC gap opening. 
The CW-type temperature dependence indicates the presence of two-dimensional (2D) AF spin fluctuations according to the SCR theory. The crossover from Fermi-liquid to CW behavior in $(T_1T)^{-1}$ correlates perfectly with the change in the resistivity results.~\cite{KasaharaBaFe2(AsP)2} As the system evolves from a Fermi liquid ($x=0.71$) towards the maximum $T_c$ ($x=0.33$) near the AF phase, the temperature dependence of the resistivity changes from $T^2$ to $T$ linear, one hallmark of NFL behavior.

We show in the next section that the CW behavior of $(T_1T)^{-1}$ is consistent with the observed temperature dependences of the electrical resistivity $\rho$ and with the predictions of a SCR model with spin-fluctuation parameters relevant to \BFAP .

\section{Analysis based on theory of spin fluctuations}
In this section, we demonstrate that experimental data of \BFAP \ and \BFCA \  are {\it quantitatively} explainable in terms of the SCR theory for two-dimensional itinerant antiferromagnets. 
All the NMR data of \BFCA \ are cited from Ref.~6.

\subsection{Outline of the self-consistent renormalization (SCR) theory}
The SCR theory gives quantitative relations between dynamical susceptibility and physical properties. 
In nearly and weakly AF metals, dynamical susceptibility above $T_{\rm N}$ for a wave vector near the AF ordering vector $\bm Q$ may generally be written as follows: 
\[\chi(\mbox{\boldmath $Q$}+\mbox{\boldmath $q$}, \omega) =\frac{\chi(\mbox{\boldmath $Q$}+\mbox{\boldmath $q$})}{1-i\omega/\Gamma_{\mbox{\boldmath $Q$}+\mbox{\boldmath $q$}}}
\]
with
\begin{eqnarray*}
[\chi(\mbox{\boldmath $Q$}+\mbox{\boldmath $q$})]^{-1} &=&[\chi(\mbox{\boldmath $Q$})]^{-1}+Aq^2, \\
\Gamma_{\mbox{\boldmath $Q$}+\mbox{\boldmath $q$}} &=&\Gamma(\kappa^2+q^2), \\ 
\kappa^2&=&1/A\chi(\mbox{\boldmath $Q$}), 
\end{eqnarray*}
where $\kappa^{-1} (\equiv \xi_{T})$ is the temperature-dependent magnetic correlation length, and $A$ and $\Gamma$ are temperature independent constants, which are the fundamental parameters of the theory.
Using above relations, the dynamical spin susceptibility $\chi ''({\bm Q}, \omega)$ is written as 
\begin{eqnarray}
\lefteqn{\chi ''(\bm Q + \bm q, \omega)  = \chi(\bm Q+\bm q)~\frac{\omega~\Gamma_{\bm Q +\bm q}}{\omega^2+\Gamma_{\bm Q +\bm q}^2}} \\ 
&=& \frac{\chi(Q)\kappa^2}{\kappa^2+q^2}\frac{\omega\Gamma(\kappa^2+q^2)}{\omega^2+[\Gamma(\kappa^2+q^2)]^2} \\
&=& \frac{\Gamma}{A}\frac{\omega}{\omega^2+[\Gamma(\kappa^2+q^2)]^2}\\ 
&=&\frac{\left(\pi\frac{\Gamma q_B^2}{2\pi}\right)}{\frac{A q_B^2}{2}}\frac{\omega}{\omega^2+\left[2\pi\frac{\Gamma q_B^2}{2\pi}\left(\frac{\kappa^2}{q_B^2}+\frac{q^2}{q_B^2}\right)\right]^2}\\ 
&=&\frac{\pi T_0}{T_A}\frac{\omega}{\omega^2 + \{2\pi T_0[y + (q/q_B)^2]\}^2},\label{chiqSCR}
\end{eqnarray}
where $q_B$ is the cut-off wave vector and has a relation of $s_0 q_B^2 / 4\pi=1$ with $s_0$ being the area per magnetic atom in the 2D plane. 
In this formula, important spin-fluctuation parameters are the following two characteristic temperatures, 
\begin{eqnarray*}
T_0 = \Gamma q_B^2/2\pi \\
T_A = A q_B^2/2, 
\end{eqnarray*}
which characterize the width of the spin excitations spectrum in frequency $\omega$ and momentum ${\bm q}$ space. 
The dimensionless inverse susceptibility $y(T)$ at ${\bm Q}={\bm Q}_{AF}$ of AF wave vectors is defined as
\begin{equation}
y(T) = \frac{\kappa^2}{q_B^2}=\frac{1}{A\chi(\bm Q)q_B^2}=\frac{1}{2T_A\chi({\bm Q})}.
\end{equation}
Here, $y_0$ is the zero temperature limit of $y$, and characterizes the proximity to the magnetic instability. 
$y_0 = 0$ indicates an AF QCP, where $\chi({\bm Q})$ diverges down to zero temperature. 

The staggered susceptibility or $y$ is determined self-consistently from the relation of the mean square local amplitude of the zero point and thermal spin fluctuations, and is calculated from 
\begin{equation}
y = y_0 + \frac{y_1t}{2} \left\{ \phi(y/t) - \phi(y/t + 1/t) \right\},
\label{eq:y} 
\end{equation}
where $t = T/T_0$, $y_1$ is the parameter which governs the mode-mode coupling of AF spin fluctuations, and $\phi(x)$ is given as,
\begin{equation}
\phi(x) = -\left(x-\frac{1}{2}\right)\log{x} + x + \log{\Gamma(x)} - \log{\sqrt{2\pi}}.
\end{equation}

\subsection{Calculations of the nuclear spin-lattice relaxation rate $T_1^{-1}$}
Nuclear spin-lattice relaxation rate $1/T_1$ is generally expressed by
\begin{equation}
\label{T1}
\frac{1}{T_1}=\frac{\gamma_N^2 T}{N_A} \lim_{\omega \rightarrow 0} \sum_{\bm q} \frac{|A_q|^2 \chi''(\bm q, \omega_0)}{\omega_0}
\end{equation} 
where $\gamma_N$ is the gyromagnetic ratio of an observed nucleus, $N_A$ is the number of magnetic atoms per unit volume, $A_q$ is the coupling constant for the hyperfine interaction between the nuclear spin and the $q$-component of the spin density, $\omega_0$ is the NMR frequency (order of milliKelvin). 
Inserting Eq.~(\ref{chiqSCR}) into Eq.~(\ref{T1}) and neglecting the $q$ dependence of $A_q$, $T_1^{-1}$ is described as follows,~\cite{MoriyaTakahashiUedaJPSJ1990}
\begin{eqnarray}
{T}_1^{-1} &=& \frac{2 \gamma^2_N A^2_{\rm hf} T}{N_A}\sum_q\frac{\pi T_0}{T_A\left[2\pi T_0\left(y+(q/q_B)^2\right)\right]^2}\\ 
                 &=& \frac{\gamma^2_N A^2_{\rm hf}}{2\pi T_{\rm A}}{\bar{T}_1^{-1}} \label{NMRSCR1}  \\
{\bar{T}_1^{-1}} &= & 2\frac{T}{T_0}\int_{0}^{1} dx \frac{x}{(y+x^2)^2} = t\left(\frac{1}{y} - \frac{1}{y+1} \right), \label{NMRSCR2}
\end{eqnarray}
where $A_{\rm hf}$ is the hyperfine coupling constant. 
Thus, $1/T_1$ directly measures the temperature dependence of $\chi({\bm Q}) = [2T_A y(T)]^{-1}$.

\subsection{Calculations of the electrical resistivity}
In the framework of the SCR theory, predominant contribution to the resistivity arises from the spin fluctuations with AF wave vectors around ${\bm Q}_{AF}$. 
The electrical resistivity in a electron system scattered by those spin fluctuations is calculated based on Boltzmann equation and is given by,~\cite{MoriyaTakahashiUedaJPSJ1990}
\begin{align}
R(T) &= r\bar{R}(T)\\ \nonumber
\bar{R}(T) &=  t\left[\phi\left(\frac{y}{t}\right) - \phi\left(\frac{(y+1)}{t}\right) \right]\\
&+ y\left[\log{\left(\frac{y}{t}\right)} - \psi\left(\frac{y}{t}\right) \right]\\ \nonumber &-(y+1)\left[\log{\left(\frac{(y+1)}{t}\right)} - \psi\left(\frac{(y+1)}{t}\right) \right]
\end{align}
where $r$ is an adjustable fitting constant which represents the coupling between the spin fluctuations and conduction electrons, and $\psi(x)$ is the digamma function.

Linear temperature dependence of the resistivity is generic to a QCP of 2D AF metals.~\cite{MoriyaTakahashiUedaJPSJ1990} 
Away from the QCP the electrical resistivity shows a crossover from the anomalous $T$ -linear dependence to the Fermi-liquid like $T^2$ behavior. 

\subsection{Analysis on spin fluctuations in \BFCA}
Inelastic neutron scattering measurements revealed that two-dimensional spin fluctuations possess the stripe correlations [$\bm Q_{\rm AF} = (0, \pi)$ or $(\pi, 0)$ in an unfolded Brillouin zone].~\cite{IshikadoPRB2011} 
The presence of the stripe correlations is also suggested from the anisotropy of NMR $1/T_1$ at the As site.~\cite{KitagawaSrFe2As2,SKitagawaAnisotropy} 
The $T_1$ anisotropic ratio $R \equiv ^{75}(1/T_1T)_{H \perp c} / ^{75}(1/T_1T)_{H \parallel c} \sim 1.5$ above $T_{\rm N}$ in BaFe$_2$As$_2$, SrFe$_2$As$_2$ and LaFeAsO is consistently understood from the anisotropic spin fluctuations due to the off-diagonal components ($B_{ac}$) of hyperfine coupling tensor $\bm B$ at the As site and from the stripe correlations of the Fe spins.~\cite{KitagawaBaFe2As2,KitagawaSrFe2As2,SKitagawaAnisotropy}
The importance of the off-diagonal terms was first pointed out by Kitagawa {\it et al.}\cite{KitagawaBaFe2As2}; the internal magnetic fields produced by diagonal terms ($B_{\alpha\alpha}$) are canceled out even if the spin correlations are stripe, since As atoms are located at the symmetrical site with respect to the four nearest neighbor Fe atoms. 
The off-diagonal terms related with the stripe correlations are discussed later. 

Eqs.~(\ref{NMRSCR1}) and (\ref{NMRSCR2}) only give the contribution of spin fluctuations around the AF wave vector ${\bm Q}_{AF}$.
Although the AF contribution around $\bm Q_{\rm AF}$ is expected to be predominant for NMR relaxation rate in \BFAP \ and \BFCA \ , there is an additional contribution arising from spin fluctuations around ${\bm q} = 0$.
The observed spin-lattice relaxation rate is thus decomposed into the following two components;
\begin{equation}
\left(\frac{1}{T_1T}\right)_{\rm obs.} = \left(\frac{1}{T_1T}\right)_{{\bm q}\sim 0} + \left(\frac{1}{T_1T}\right)_{{\bm q}\sim {\bm Q}_{AF}}. 
\end{equation}

\begin{figure}[tb]
\begin{center}
\includegraphics[width=8.5cm]{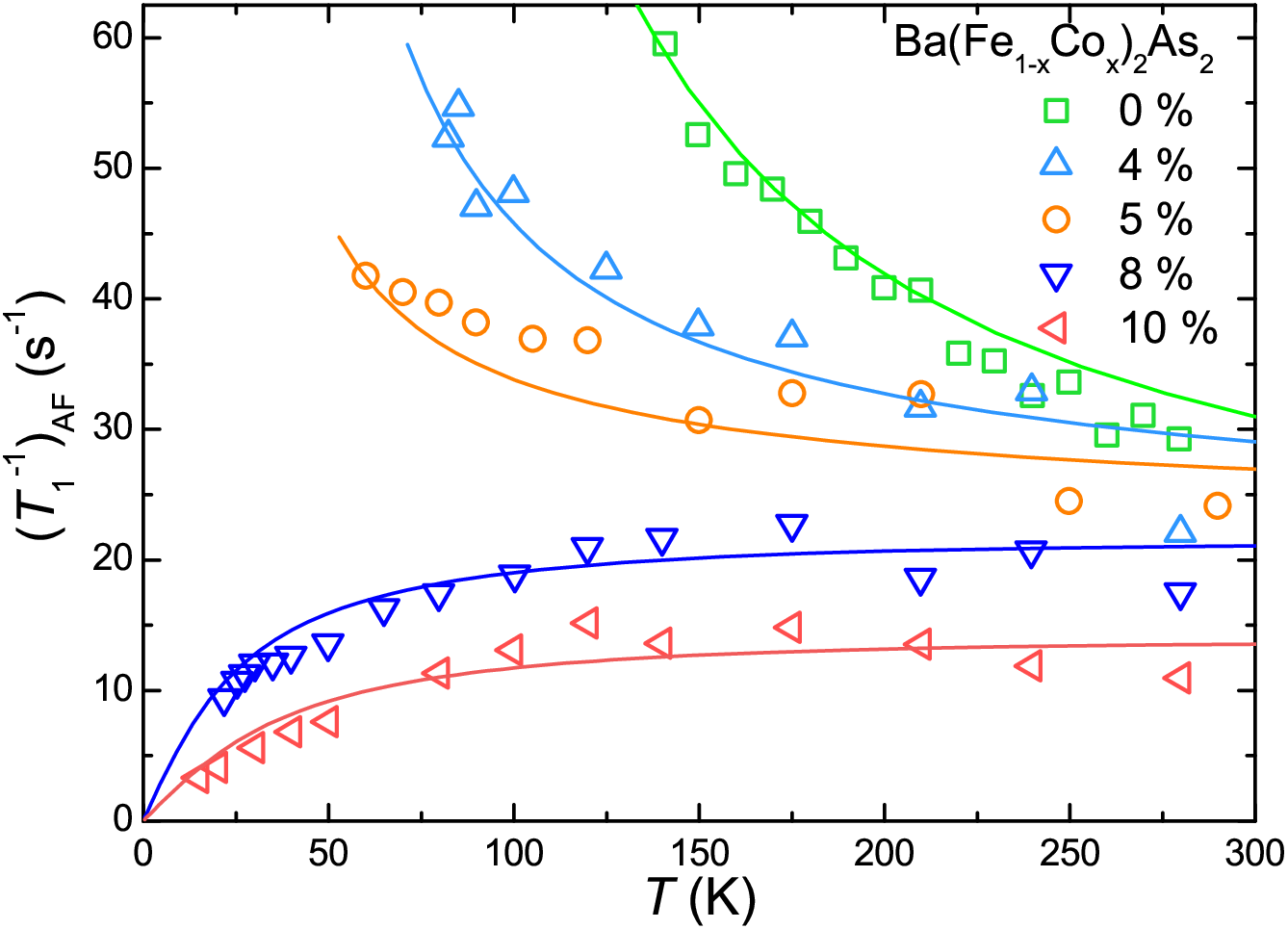}
\end{center}
\caption{(Color online) The $T$-dependence of the NMR relaxation rate arising from the ${\bf q}\sim {\bf Q}_{\rm AF}$ mode of spin fluctuations $(T_1)^{-1}_{\rm AF}$ of Ba(Fe$_{1-x}$Co$_{x}$)$_2$As$_2$ for $H \parallel ab$, cited from Ref.~6. The solid lines represent simulations with the SCR parameters listed in Table~\ref{tableBFCA}.}
\label{BFCAT1}
\end{figure}

The experimental NMR results of \BFCA \ reported by Ning {\it et al.}~\cite{NingPRL2010} show significant AF fluctuations near the optimal doping of $x\sim0.06$, and that the AF spin fluctuations are systematically suppressed by Co doping. 
In addition, they reported that $(T_1T)^{-1}$ decreases on cooling for over-doped samples, as observed in LaFeAs(O,F),~\cite{NakaiJPSJ2008,NakaiNJP2009,SKitagawaAnisotropy} indicating that AF spin stripe correlations are not significant in highly over-doped samples. 
These results are in good agreement with the inelastic neutron scattering measurements.~\cite{MatanPRB2010} 
Since the NMR results of \BFCA \ indicate that the background term of $(T_1T)^{-1}$, which is ascribable to $(T_1T)^{-1}_{{\bf q}\sim 0}$, shows non-monotonic behavior, the analysis of contributions from AF fluctuations is less straightforward than \BFAP, which will be shown below. 
By assuming the temperature dependence of the background term of $(T_1T)^{-1}$ is identical with that of the Knight shift, 
Ning {\it et al.} estimated the AF contribution and found that its temperature dependence follows Curie-Weiss-type $(T_1T)^{-1}_{{\bm q}\sim {\bm Q}_{AF}} = C/(T + \theta)$ as observed in \BFAP.~\cite{AhilanPhysicaC} 
They thus employed the following phenomenological two-component model:
\begin{align}
(T_1T)^{-1}_{\rm obs.} &= (T_1T)^{-1}_{{\bm q}\sim {\bm Q}_{AF}} + (T_1T)^{-1}_{{\bm q}\sim 0} \\ 
(T_1T)^{-1}_{{\bm q}\sim 0} &= \alpha K_{\rm spin} = \alpha\left(a + b\exp{(-\Delta/k_BT)}\right). 
\end{align}
By using their estimation of $(T_1T)^{-1}_{{\bm q}\sim 0}$,~\cite{NingPRL2010,AhilanPhysicaC} the doping dependence of $(T_1)^{-1}_{{\bm q}\sim {\bm Q}_{AF}}$ was obtained as shown in Fig.~\ref{BFCAT1}.  
Since $(T_1)^{-1} = const.$ behavior is an indication of the verge of a 2D AF QCP (see the next section E), we expect a critical Co concentration of $0.05< x <0.08$ in \BFCA.

\begin{figure}[tb]
\begin{center}
\includegraphics[width=7.5cm]{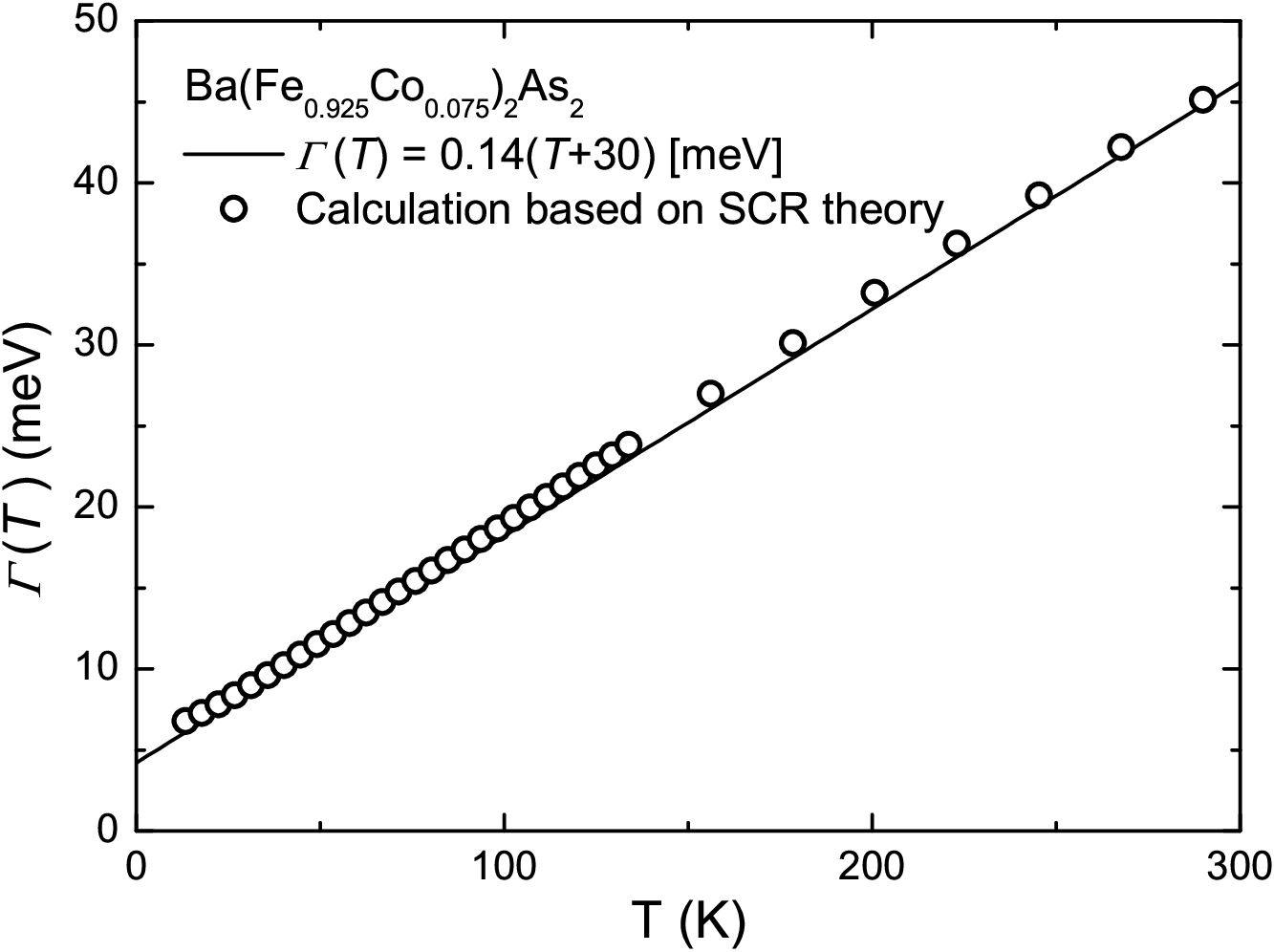}
\end{center}
\caption{The experimental\cite{InosovNaturePhys} (solid line) and calculated (circles) damping constant $\Gamma(T)$ for the dynamical susceptibility $\chi"(Q,\omega) = \frac{\chi_T\Gamma(T)\omega}{\omega^2 + \Gamma^2(T)(1 + \xi_T^2|\bm{Q}-\bm{Q}_{AFM}|)}$ observed in Ba(Fe$_{0.925}$Co$_{0.075}$)$_2$As$_2$. The circles represent a calculation with $y_0= 0.025$, $y_1=2.5$, and $T_0=450$ K.}
\label{Neutron}
\end{figure}

For simulating the NMR data, we need to determine $y_0$, $y_1$, $T_0$, and $T_A$.
In order to narrow down SCR parameters, we analyzed inelastic neutron scattering data.~\cite{InosovNaturePhys,MatanPRB2010} 
Inosov {\it et al.} reported that the temperature dependence of the damping constant $\Gamma(T)$ of the dynamical spin susceptibility for nearly optimally doped Ba(Fe$_{0.925}$Co$_{0.075}$)$_2$As$_2$ ($T_c = 25$ K) shows a linear temperature dependence $\Gamma(T) = 0.14(T+30) [{\rm meV}]$.
The $\Gamma(T)$ can be calculated in the SCR theory as follows,~\cite{MoriyaTakimotoJPSJ,KambePRB2008USn3}
\begin{equation}
\Gamma(T) = 2\pi T_0y(T).
\end{equation}
The $T$-dependence of the $\Gamma$ is thus sensitive to the parameters of $y_0$, $y_1$, and $T_0$. 
In this way, we simulate the NMR relaxation rate and neutron scattering data of $x = 0.08$ as shown in Fig.~\ref{BFCAT1} and ~\ref{Neutron}, indicating very good agreement with the experiments.

In order to determine the SCR parameters of other Co doping, we used the data of magnetic susceptibility and specific heat as follows.
In the framework of the SCR theory, we can relate the magnetic susceptibility to $T_A$ using the following relation;~\cite{MoriyaTakimotoJPSJ}
\begin{equation}
T_A = 0.75 / \chi \mbox{(in emu/mol)}.
\label{TAchi}
\end{equation}
The estimated $T_A$ of $x=0.08$ from the NMR and neutron scattering data corresponds to the susceptibility at $T \simeq 200$ K.~\cite{WangNJP2009} 
For other $x$ values, we thus use the $T_A$ values estimated from the susceptibility at 200 K for our calculation.~\cite{WangNJP2009} 
The doping dependence of $T_0$ is estimated from the reported specific heat experiment~\cite{HardyCoDopingDepBa122} by using the following relation;~\cite{MoriyaUedaJPSJ1994,MoriyaTakimotoJPSJ} 
\begin{equation}
\gamma = \frac{6200}{T_0}(2x_c - \pi y_0^{-1/2})  \hspace{0.5cm}    \mbox{[mJ/mol~K$^{2}$]}.
\label{specific}
\end{equation}
where $x_c$ is the cut-off wave vector of which magnitude is the order of unity. 
Note that we used the first term for $y_0 < 0$.~\cite{MoriyaTakimotoJPSJ}

\begin{figure}[tb]
\begin{center}
\includegraphics[width=7.5cm]{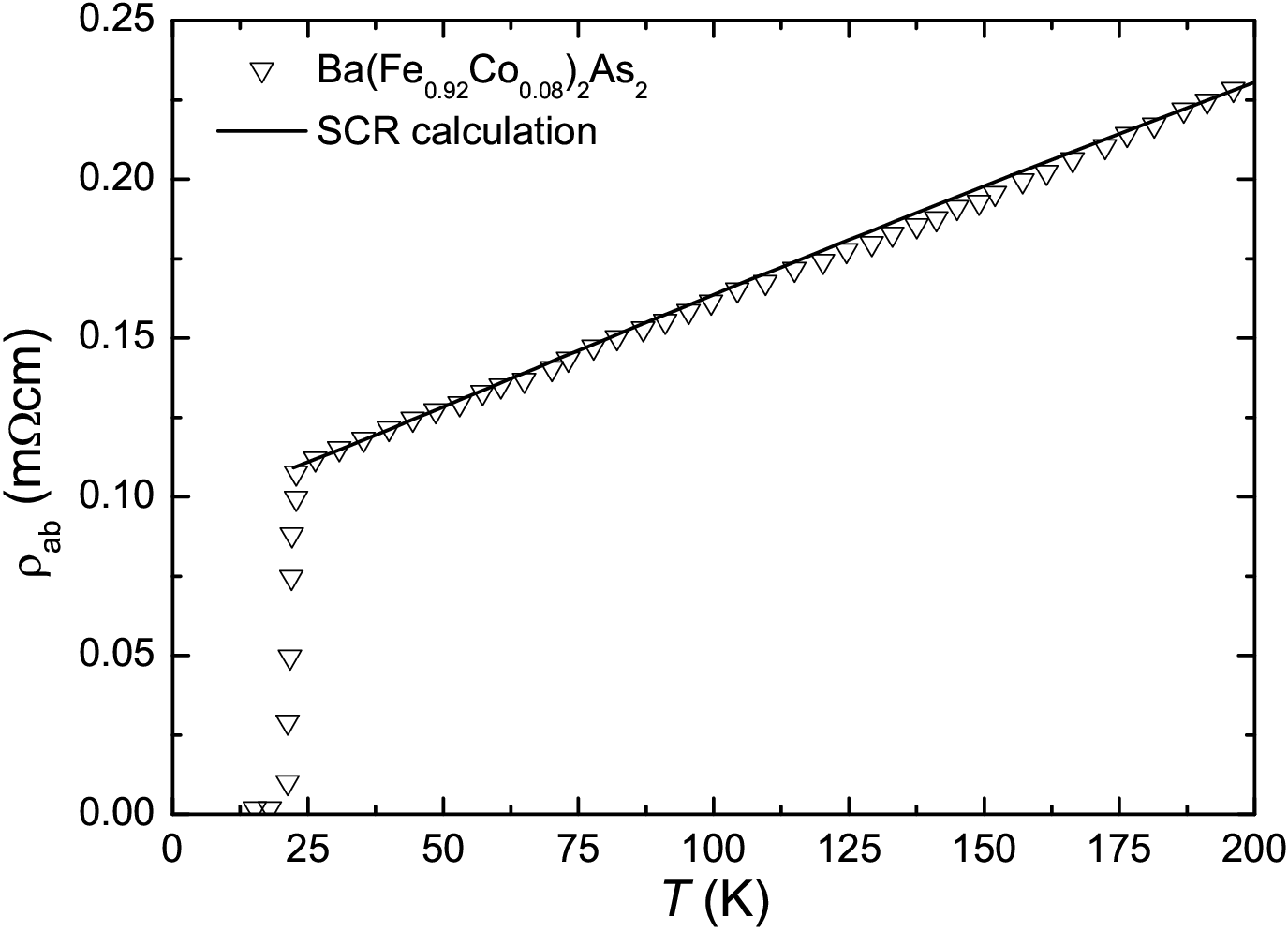}
\end{center}
\caption{The $T$-dependence of the electrical resistivity $\rho_{ab}$ of Ba(Fe$_{0.92}$Co$_{0.08}$)$_2$As$_2$, cited from Ref.~27. The solid line represents a calculation using $y_0=0.025$, $y_1=2.5$, and $T_0 = 450$ K, and residual resistivity $\rho^{ab}_0=0.1$ m$\Omega$cm.}
\label{BFCArho}
\end{figure}

In this way, we simulate the Co concentration dependence of the NMR relaxation rate as shown in Fig.~\ref{BFCAT1} with the SCR parameters as listed in Table~\ref{tableBFCA}, indicating very good agreement with the experimental data. 
Using the same parameter, we also calculated the temperature dependence of the electrical resistivity and found good agreement with the experimental result as shown in Fig.~\ref{BFCArho}. 

\begin{table}[h]
\begin{center}
\begin{tabular}{r|r|r|r|r} 
 & $y_0$ & $y_1$ & $T_0$ (K) & $T_A$ (K) \\\hline\hline
$x$ = 0.0 & -0.4 & 8.0 & 460 & 800 \\
0.04 & -0.07 & 3.0 & 510 & 950 \\
0.05 & -0.03 & 2.7 & 530 & 960 \\
0.08 & 0.025 & 2.5 & 450 & 1050 \\
0.1 & 0.05 & 4.0 & 420 & 1200 \\
\end{tabular}
\end{center}
\caption{The SCR parameters of \BFCA.}
\label{tableBFCA}
\end{table}

\subsection{Analysis on spin fluctuations in \BFAP}
Following the similar procedure as in \BFCA, we here analyze the experimental NMR results of \BFAP.
Because the uniform susceptibility of \BFAP \ is almost temperature independent, the NMR relaxation rate arising from the small-${\bm q}$ fluctuations $(T_1T)^{-1}_{{\bm q}\sim 0}$ is also expected to be temperature independent. 
In order to estimate $(T_1T)^{-1}_{{\bm q}\sim 0}$, the observed $(T_1T)^{-1}$ is fit by the CW-type equation: $(T_1T)^{-1} = a +b/(T+\theta)$. 
We relate the first constant term $a$ with the $(T_1T)^{-1}_{{\bm q}\sim 0}$ and estimated $(T_1)^{-1}_{{\bm q}\sim {\bm Q}_{AF}}$ as shown in Fig.~\ref{BFAPSCRT1}. 
The nearly constant $(T_1)^{-1}_{{\bm q}\sim {\bm Q}_{AF}}$ of $x=0.33$ suggests that $y_0$ is very close to zero at $x=0.33$.~\cite{MoriyaTakahashiUedaJPSJ1990} 

\begin{figure}[tb]
\begin{center}
\includegraphics[width=8cm]{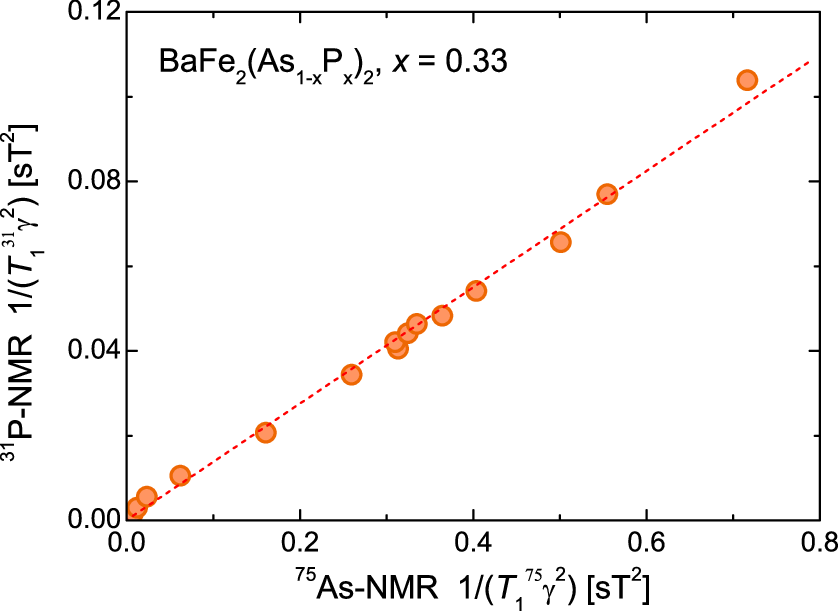}
\end{center}
\caption{(Color online) The $T_1^{-1}$ of $^{31}$P and $^{75}$As in \BFAP \  with $x=0.33$ is plotted. We estimated the hyperfine coupling constant of $^{31}$P nucleus from the slope as $^{31}A_{\rm hf}$=0.674 T/$\mu_B$ by using $^{75}$A$_{\rm hf} = $ 1.72 T/$\mu_B$.~\cite{KitagawaBaFe2As2}}
\label{PvsAs}
\end{figure}

\begin{figure}[tb]
\begin{center}
\includegraphics[width=8cm]{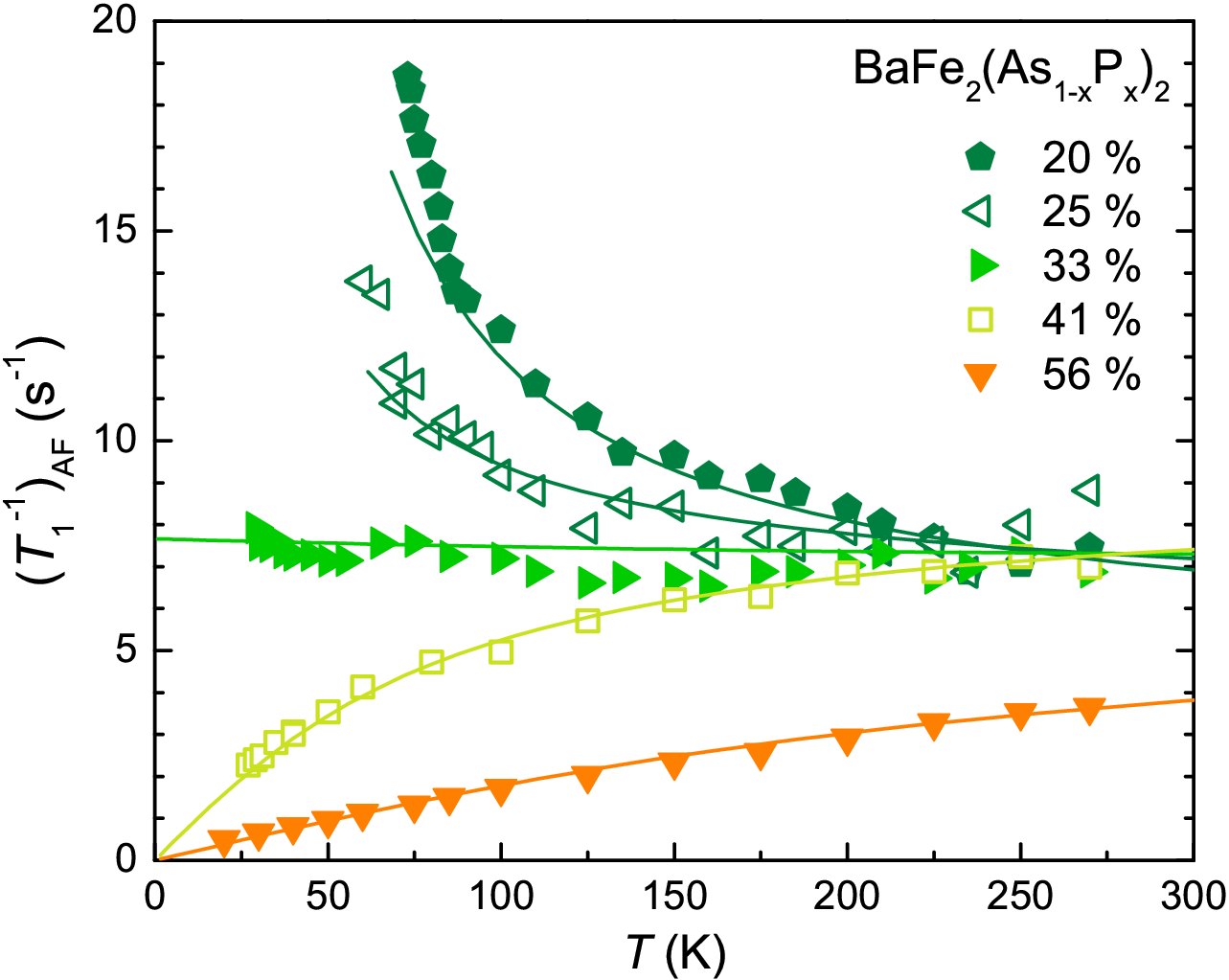}
\end{center}
\caption{(Color online) The experimental and calculated NMR relaxation rate arising from ${\bm q}\sim {\bm Q}_{AF}$ mode of spin fluctuations in \BFAP. The data points represent the experimental data, whilst the solid lines indicate the calculated data using the SCR parameters listed in Table~\ref{tableBFAP}.}
\label{BFAPSCRT1}
\end{figure}

For simulations of $T_1^{-1}$, one needs the hyperfine coupling constant $A_{\rm hf}$ of $^{31}$P nucleus. 
In order to estimate it, it is reasonably assumed that the $T_1^{-1}$ of $^{31}$P is determined by the off-diagonal terms of the hyperfine coupling tensor, as is the case for the $T_1^{-1}$ of $^{75}$As in BaFe$_2$As$_2$.~\cite{KitagawaBaFe2As2}
Figure~\ref{PvsAs} displays the $T_1^{-1}$ of $^{75}$As for $x = 0.33$ plotted against that of $^{31}$P with temperature as an implicit parameter. 
Since the $T_1^{-1}$ of $^{31}$P is proportional to that of $^{75}$As as shown in Fig.~\ref{PvsAs}, we can estimate $^{31}A_{\rm hf}^{\rm off} = $6.37 kOe/$\mu_B [\equiv 4 (^{31}B_{ac})]$ for $^{31}$P nucleus by using $^{75}A_{\rm hf}^{\rm off} = $17.2 kOe/$\mu_B [\equiv 4 (^{75}B_{ac})]$ for $^{75}$As nucleus.~\cite{KitagawaBaFe2As2} 
We also assume in our calculations that the hyperfine coupling constant is independent of P concentration.

Although complete P concentration dependence of magnetic susceptibility and specific heat is not reported in \BFAP, we estimate $T_0$ and $T_A$ as follows. 
The characteristic spin fluctuation energy $T_0$ of $x=0.33$ with $y_0 = 0$ can be estimated from reported specific heat experiments~\cite{KimSpecificHeatBaFe2(AsP)2} by using Eq.~(\ref{specific}). 
Assuming that the P concentration dependence of $\gamma$ is identical with that of $K_{\rm spin}$ which is the measure of the DOS at the Fermi energy,~\cite{NakaiPRL2010} 
we can estimate $\gamma$ for other P concentrations and obtain the P concentration dependence of $T_0$ using Eq.~(\ref{specific}). 
We here neglect the second term in Eq.~(\ref{specific}) for simplicity. 
In order to estimate $T_A$ from Eq.~(\ref{TAchi}), we assume the magnetic susceptibility $\chi$ is proportional to $K_{\rm spin}$. 
By using $\chi = 9.4\times10^{-4}$ emu/mol at 200 K and $\gamma = 27$ mJ/molK$^2$ in BaFe$_2$As$_2$,~\cite{WangNJP2009,HardyCoDopingDepBa122} we thus estimate the P concentration dependence of $T_A$.

\begin{figure}[tb]
\begin{center}
\includegraphics[width=8cm]{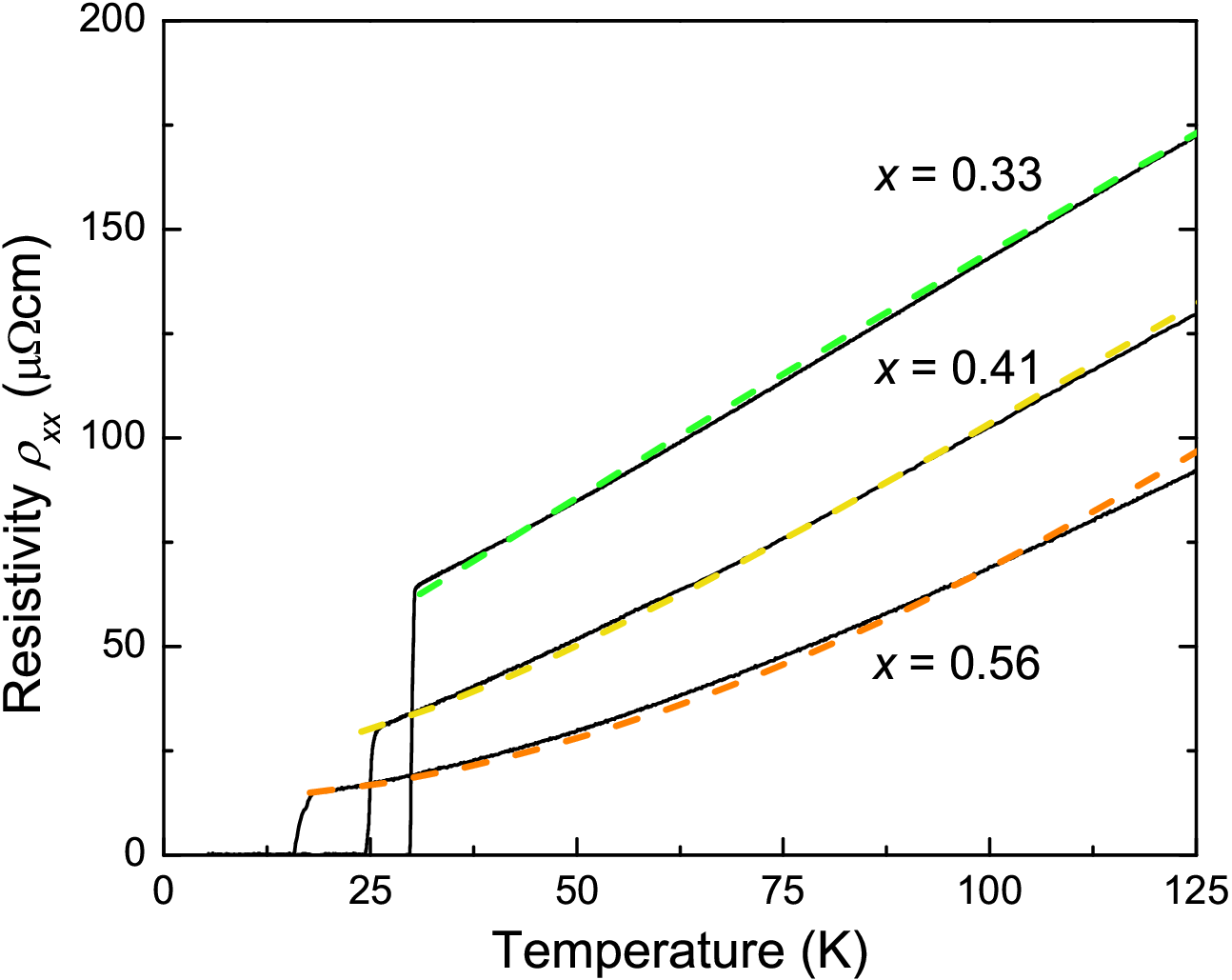}
\end{center}
\caption{(Color online) The experimental~\cite{KasaharaBaFe2(AsP)2} and calculated electrical resistivity of \BFAP. 
The solid lines represent the experimental data, whilst the dotted lines indicate the calculated data using the SCR parameters listed in Table~\ref{tableBFAP}.}
\label{BFAPrho}
\end{figure}

By using the SCR parameters listed in Table \ref{tableBFAP}, which are obtained from our NMR simulation shown in Fig.~7, we calculated the temperature exponent of electrical resistivity. 
For 2D AF fluctuations, a $T$-linear resistivity is expected near the QCP.~\cite{MoriyaUedaRepProgPhys} 
Away from the QCP, the temperature dependence of the resistivity crossovers to a Fermi-liquid-like $T^2$ as $T$ decreases. 
The experimental data is actually consistent with the simulated temperature dependence as shown in Fig.~\ref{BFAPrho}.

According to the SCR theory, we can also estimate the in-plane spin correlation length $\xi(T)$ and the damping constant $\Gamma(T)$ from $(\sqrt{4\pi y})^{-1}$ and from $2\pi T_0 y$, respectively.~\cite{MoriyaUedaAdvPhys} 
We calculated $\xi/a$ and $\Gamma(T)$ at $T_c$ for different P concentration as shown in Table~\ref{tableBFAP}, which may be confirmed by future neutron scattering experiments. 
\begin{table}[htbp]
\begin{center}
\begin{tabular}{r|r|r|r|r|r|r} 
 & $y_0$ & $y_1$ & $T_0$ (K) & $T_A$ (K)& $\xi(T_c)/a$ & $\Gamma(T_c)$ (meV) \\\hline\hline
$x$ = 0.0 & -0.4 & 8.0 & 460 & 800 &  \\
0.2 & -0.15 & 15 & 760 & 1320 & \\
0.25 & -0.05 & 10 & 770 & 1340 & 3.0 & 3.7 \\
0.33 & 0 & 8.0 & 780 & 1350 & 1.9 & 9.2 \\
0.41 & 0.06 & 5.0 & 800 & 1390 & 1.1 & 27 \\
0.56 & 0.2 & 6.0 & 850 & 1480 & 0.6 & 96 \\
\end{tabular}
\end{center}
\caption{The SCR parameters of \BFAP. The in-plane spin correlation length $\xi/a$ and damping constant $\Gamma$ at $T_c$ are also shown. Note that $a$ is the in-plane lattice constant.}
\label{tableBFAP}
\end{table}

\section{Discussion}
\subsection{Phase diagrams}
\begin{figure}[tb]
\begin{center}
\includegraphics[width=8.5cm]{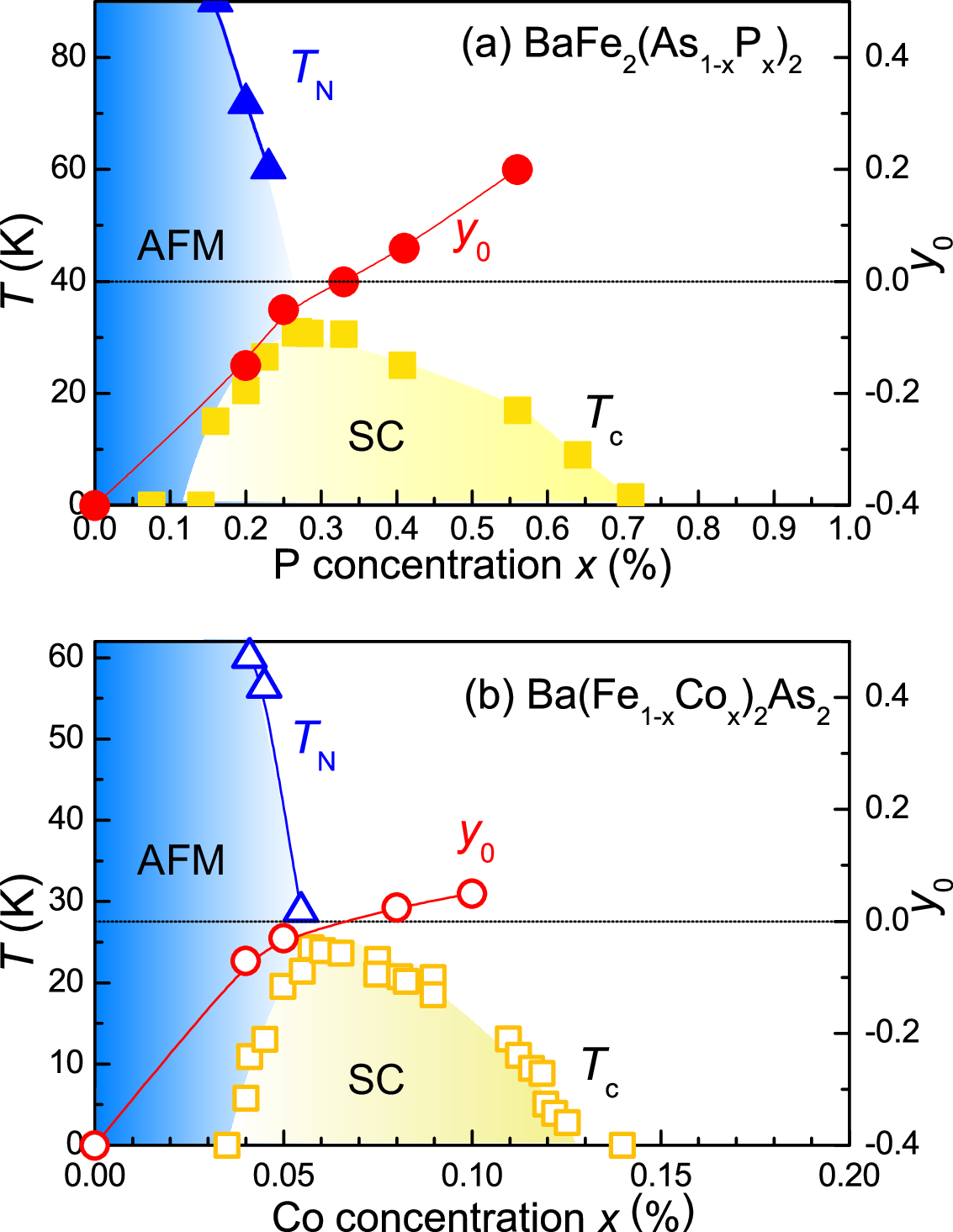}
\end{center}
\caption{(Color online) Phase diagrams of \BFAP \  and \BFCA, where $T_N$, $T_c$ denotes an AF transition temperature, SC transition temperature, respectively.
In the both materials, the concentration where $T_c$ peaks ($x\sim0.3$ for \BFAP, and $x\sim0.06$ for \BFCA) exists near the region where $y_0=0$. 
Since $y_0=0$ corresponds to an AF QCP, these phase diagrams suggest a close link between superconductivity and AF quantum criticality.}
\label{PhaseDiagram}
\end{figure}

The phase diagrams of \BFAP \  and \BFCA \  are plotted in Fig~\ref{PhaseDiagram}.
The $y_0$ increases with chemical substitution from a negative value in BaFe$_2$As$_2$ to nearly zero around an optimal concentration; $x\sim$0.3 for \BFAP \ and $x\sim0.06$ for \BFCA. 
Since $y_0$ is a measure of the closeness to a QCP, this indicates that their optimal concentration corresponds to an AF QCP and the closeness to the QCP is controllable by P and Co substitution. 

\subsection{Spin fluctuation temperature $T_0$, $T_A$ versus SC transition temperature $T_c$}
\begin{figure}[tb]
\begin{center}
\includegraphics[width=8.5cm]{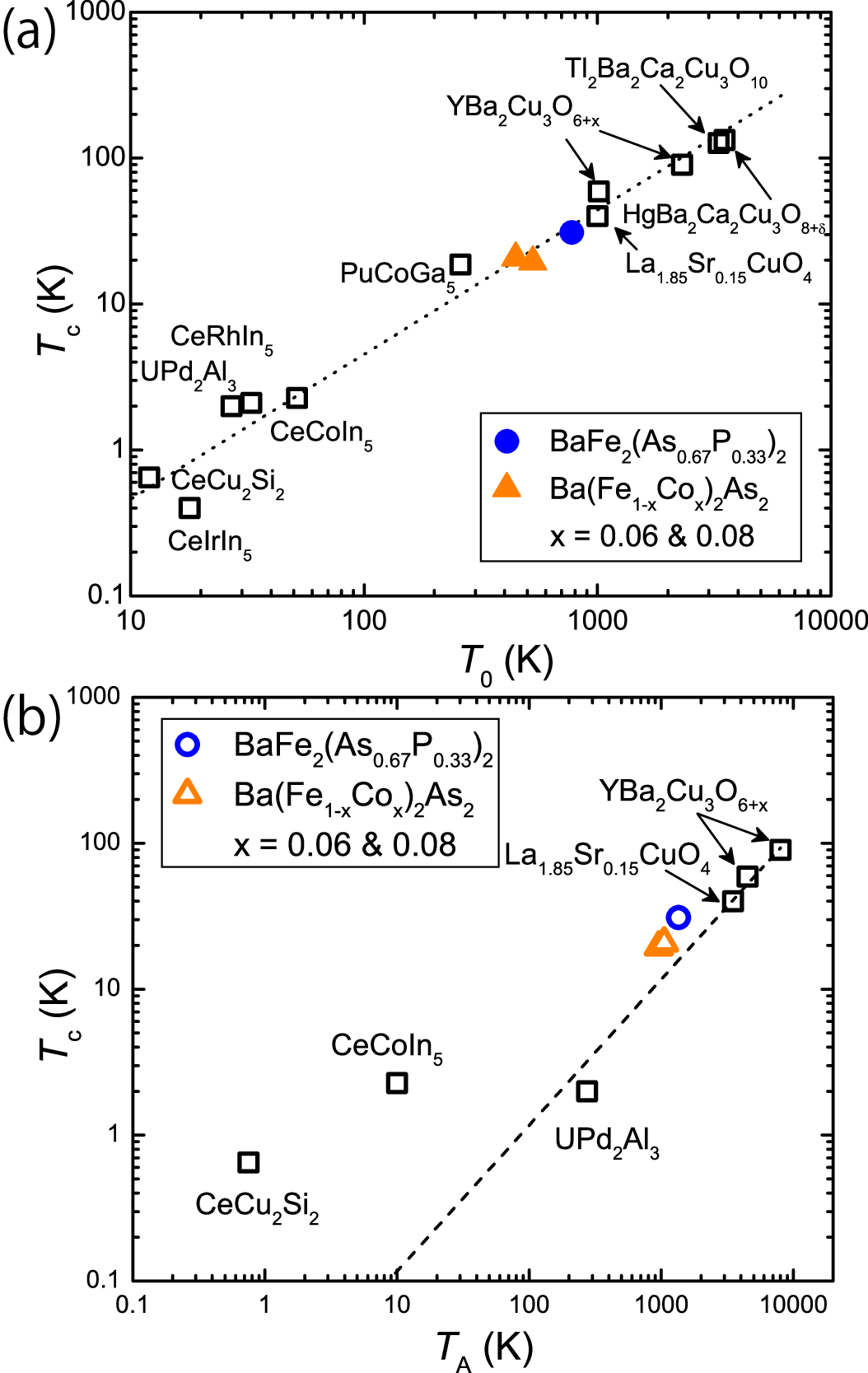}
\end{center}
\caption{(Color online) The SC transition temperature $T_c$ vs. the spin fluctuation temperature (a) $T_0$ and (b) $T_A$ for various superconductors. The data for $T_c$ and $T_0$ were cited from Refs.~35 and 37. The dotted lines represent linear curve fittings. 
$T_c$ is linear with $T_0$ as suggested in Ref.~35, and the iron-pnictide superconductors lie on the same line. 
This may be the signature that Fe pnictides, heavy fermion, and cuprate superconductors are mediated by AF spin fluctuations. }
\label{TcVST0}
\end{figure}
In \BFAP \  and \BFCA \ , the SC phase exists next to the AF phase, and $T_c$ is maximum nearly at an AF QCP, i.e. $y_0 \sim 0$ as shown in Fig.~\ref{PhaseDiagram}. 
This strongly suggests that there is an intimate link between superconductivity and antiferromagnetism in iron-pnictide superconductors. 
This is reminiscent of heavy-fermion (HF) superconductors, particularly Ce-based superconductors such as CeCu$_2$Si$_2$ and Ce$M$In$_5$ ($M$: Co, Rh, and Ir).~\cite{PfleidererReview} 
In these Ce-based HF superconductors, superconductivity occurs near an AF QCP, which is induced by competition between the  Ruderman-Kittel-Kasuya-Yosida (RKKY) interaction and the Kondo effect. 
A number of experiments in these HF superconductors reported NFL behavior (e.g. $\rho(T) \propto T^{\alpha}$ over a wide $T$ range at low temperatures) near the QCP. 
The NFL behavior is ascribed to AF spin fluctuations with a quantum critical nature, and the AF spin fluctuations likely induce unconventional superconductivity with a $d$-wave order parameter.
Similarly, AF fluctuations are also suggested for a likely candidate of the pairing mechanism for high-$T_c$ cuprate superconductors where significant NFL behavior is observed, 
although understanding of the pseudogap behavior in the normal state has not been settled.
  
In order to understand a relationship between AF spin fluctuations and superconductivity, we plot SC $T_c$ against spin-fluctuation parameters, $T_0$ and $T_A$ of \BFAP \  and \BFCA \ as well as those of unconventional superconductors in Fig.~\ref{TcVST0}. 
Note that only optimal \BFAP \  and nearly optimal \BFCA \ are plotted, since the (nearly) optimal samples are close to an AF QCP.~\cite{MoriyaUedaRepProgPhys,CurroNature05} 
The linear scaling between the spin fluctuation temperature $T_0$ and $T_c$ in Ce-based HF superconductors and the cuprates was interpreted as an indication of spin-fluctuation mediated superconductivity in these unconventional superconductors and that a higher spin fluctuation temperature can give rise to a pairing interaction and thus resulting in higher $T_c$.~\cite{MoriyaUedaRepProgPhys,CurroNature05} 
Interestingly, optimal \BFAP \  has a higher $T_0$ and $T_c$ than \BFCA, and these ``122'' iron-pnictide superconductors have intermediate values of $T_0$ and $T_c$ among other unconventional superconductors. 
This suggests that the physics of ``122'' iron-pnictide superconductors may be more closely related to the physics of HF and cuprate superconductors than previously expected, and they may be classified into magnetically mediated superconductors. 
Moreover, only the optimal superconductivity lies on the curve. 
This suggests that quantum criticality is another important ingredient for understanding of the linearity between spin fluctuation temperature and $T_c$, as discussed in the HF superconductors such as Ce$M$In$_5$.~\cite{GegenwartNatPhys,ParkNature2006}

\begin{figure}[tb]
\begin{center}
\includegraphics[width=8cm]{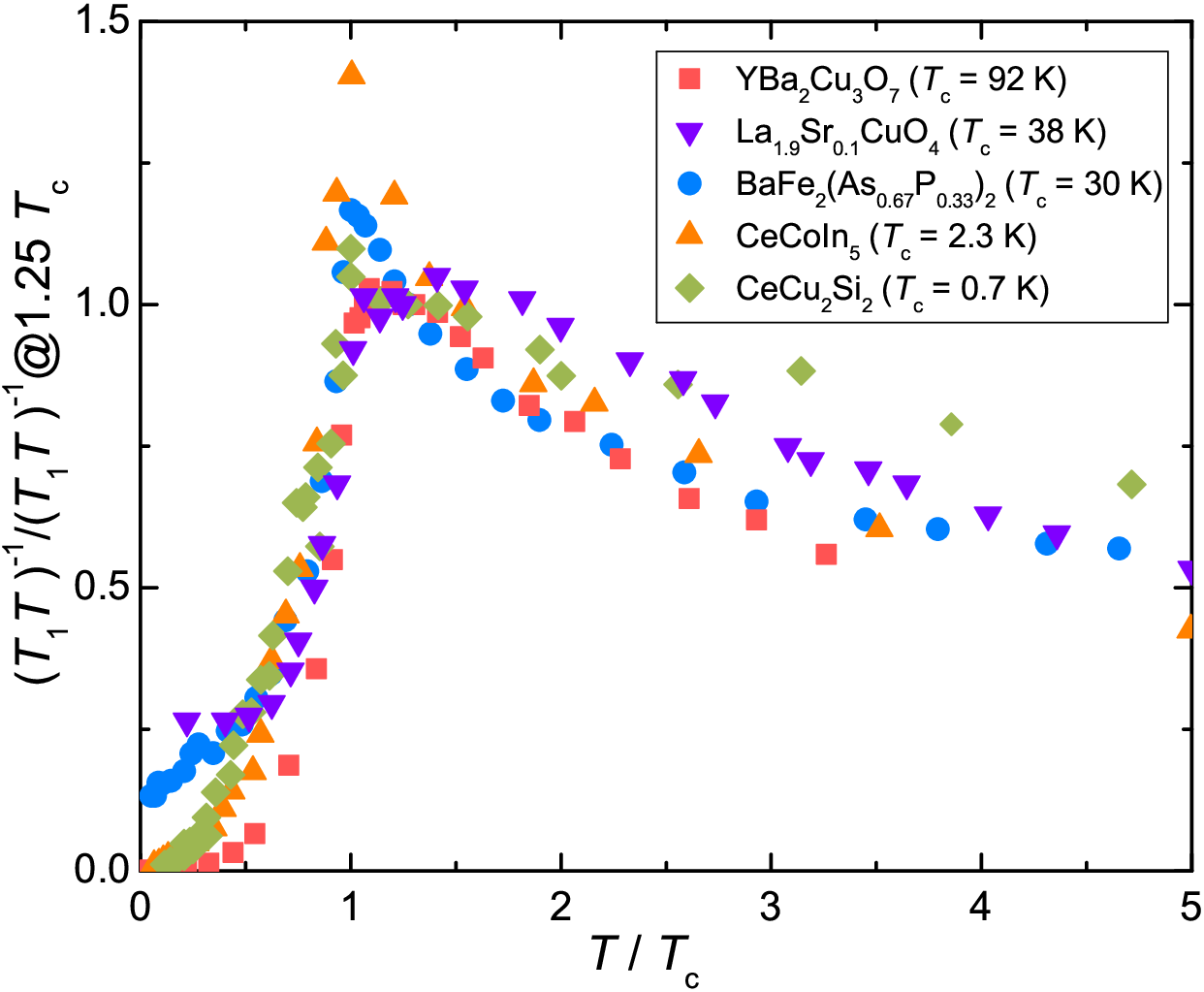}
\end{center}
\caption{(Color online) The $T$-dependence of $(T_1T)^{-1}$ normalized by $(T_1T)^{-1}$ at 1.25$T_c$ in various unconventional superconductors~\cite{IshidaYBCO,OhsugiLSCO,KawasakiCeCoIn5,IshidaCeCu2Si2} are plotted against $T/T_c$. $(T_1T)^{-1}$ at 1.25$T_c$ is adopted in order to avoid the suppression by the pseudogap effect. The characteristic energy of the spin fluctuations in these compounds seems to be scaled to $T_c$, since the normalized $(T_1T)^{-1}$ data are approximately on the same curve.}
\label{variousT1T}
\end{figure}

In addition, it is noteworthy that $T_A$ is roughly scaled to $T_c$ as shown in Fig.~\ref{TcVST0} (b), and hence to $T_0$. 
This implies that the spin-fluctuation spectra are renormalized with $T_0$ as indeed inferred from Fig.~\ref{variousT1T}, where the renormalized $(T_1T)^{-1}$ of various unconventional superconductors approximately scales onto a same curve against $T/T_c$. 
Because $T_c$ values in these superconductors are different by two orders of magnitude, the scaling of $(T_1T)^{-1}$ is surprising and suggests that spin-fluctuation spectra are related with their unconventional superconductivity. 
   
\begin{figure}[tb]
\begin{center}
\includegraphics[width=8cm]{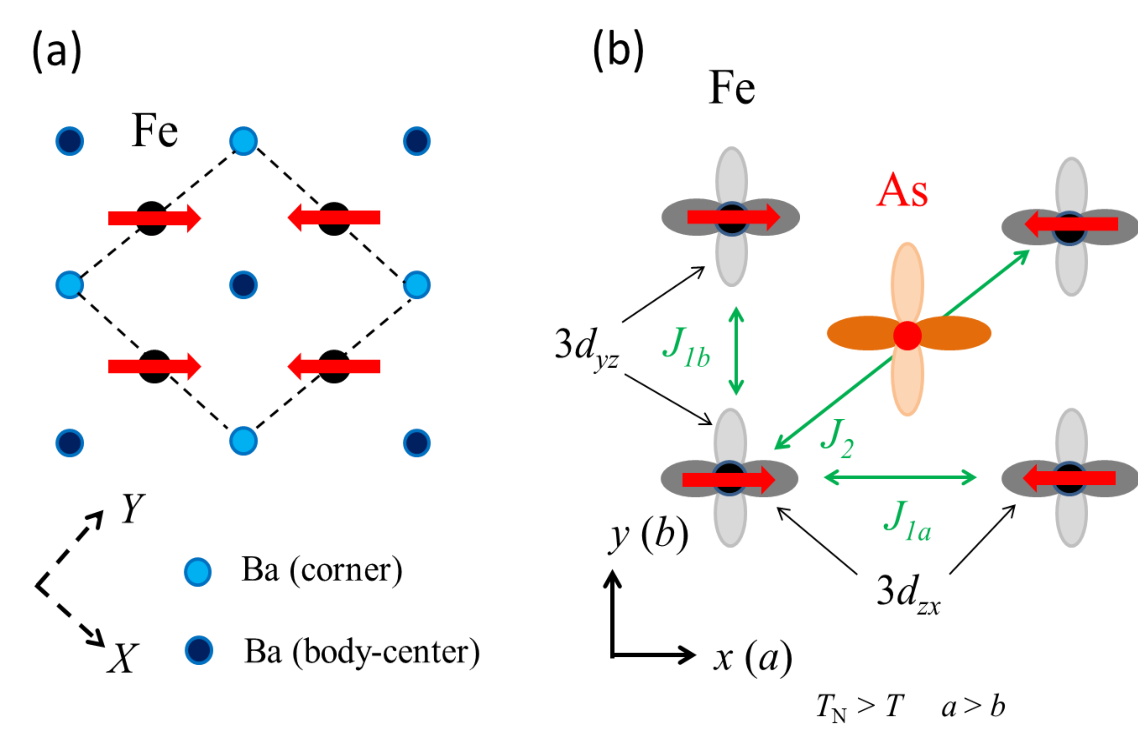}
\end{center}
\caption{(Color online) (a) Low-temperature magnetic structure at the Fe layer. The Ba sites are also shown. The dotted lines indicate the tetragonal unit cell above $T_{\rm S}$, and are deformed below $T_{\rm S}$. The stripe magnetic structure is shown by the arrows. (b) As $4p_{x, y}$ and Fe $3d_{yz}$ and $3d_{zx}$ orbitals are shown. The difference of the electronic population is shown by contrasting density. Magnetic interactions between nearest neighbor, next nearest neighbor are denoted by arrows. }
\label{FeAs}
\end{figure}
\subsection{Coupling between AF spin fluctuations and lattice instability}
In this section, we comment on the relationship between magnetism and lattice structure in the ``122'' compounds. 

A structural transition from the high-temperature tetragonal to low-temperature orthorhombic phases occurs at $T_{\rm S}$ that is identical to $T_{\rm N }$ or just above $T_{\rm N}$. 
Since the structural unit vectors rotates by 45 degree at the transition, the dotted lines in Fig.~\ref{FeAs} represent the distorted basal plane below $T_{\rm S}$, which is a unit cell above $T_{\rm S}$. 
An unusual anisotropic interaction ($J_{1a} >J_{1b}, J_{1a} \sim J_{2}$) was reported in the ordered state from the neutron scattering experiments.~\cite{JohnstonReview} 
We suggest that the anisotropic interactions are reasonably understood by the coupling between four Fe sites by way of the As site as follows. 
Kitagawa {\it et al.} reported that the electric quadrupole interaction ($\nu$) at the As site changes significantly below $T_{\rm S}$:  $\nu_a$ along the $a$ axis becomes largest, although the difference between the lattice constant $a$ and $b$ is less than 1 \%.~\cite{KitagawaBaFe2As2} 
This strongly suggests that the isotropic charge distribution above $T_{\rm S}$ becomes anisotropic, resulting in a higher electron occupation in $4p_x$ than that in $4p_y$. 
A similar conclusion was drawn from ARPES experiments.~\cite{ShimojimaPRL} 
Such an imbalance of occupation implies that degenerate Fe 3$d_{xz}$ and 3$d_{yz}$ orbitals are lifted due to nonequivalent mixing with As 4$p_x$ and 4$p_y$ orbitals, in other words, orbital ordering of the Fe 3$d$ orbitals is realized. This symmetry breaking naturally leads to a deviation of the exchange interaction $J_1$ between the nearest-neighbor Fe spins. The corresponding orthorhombic distortion can make a $J_{1b}$ ferromagnetic interaction rather than an antiferromagnetic one, following the Goodenough-Kanamori rules, because the Fe-As-Fe bond angle for $J_{1a}$ becomes close to 90 degrees. Such a tendency is consistent with the recent studies by the neutron scattering measurements.~\cite{DaiNaturePhys2012} 

\begin{figure}[tb]
\begin{center}
\includegraphics[width=8cm]{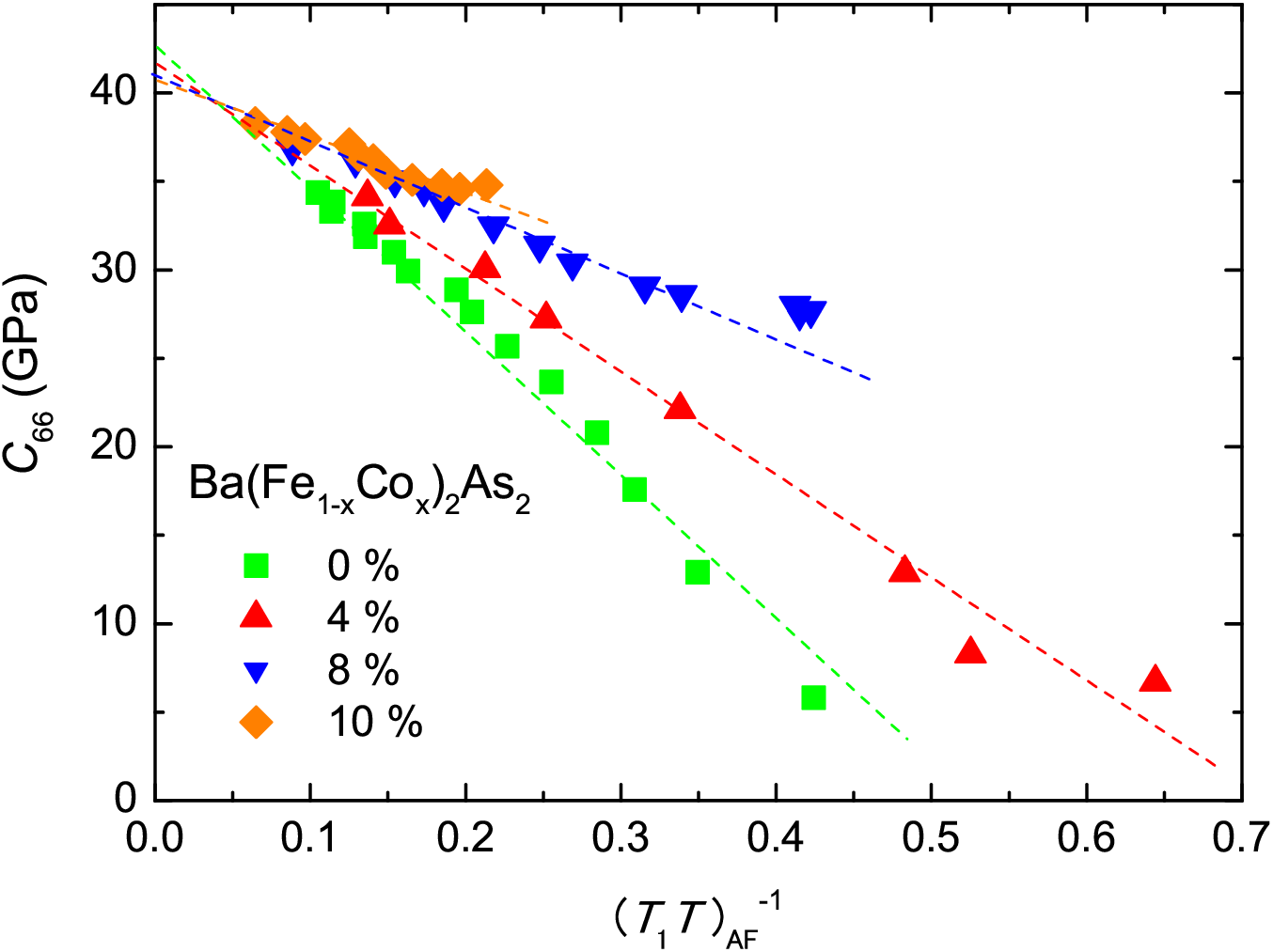}
\end{center}
\caption{(Color online) The $T$-dependence of the elastic constant $C_{66}$ in \BFCA, measured by Yoshizawa {\it et al.}~\cite{YoshizawaJPSJ2012}, is plotted against $(T_1T)^{-1}_{\rm AF}$ of Fig. 6. 
Apparent proportionality holds between $C_{66}$ and $(T_1T)^{-1}_{\rm AF}$. 
The dotted lines are guide to the eyes.
}
\label{C66vsT1T}
\end{figure}
It is naturally expected that the above anisotropic correlations persist well above $T_{\rm N}$, since the stripe AFM fluctuations are observed in the tetragonal phase, and thus the orbital fluctuations linked with the characteristic stripe AF spin correlations are anticipated above $T_{\rm N}$.
Actually, such lattice dynamics was observed with ultrasonic experiments.
Goto {\it et al.} and Yoshizawa {\it et al.} reported independently that the elastic constant $C_{66}$ in \BFCA \ shows a large elastic softening towards $T_{\rm S}$.~\cite{GotoJPSJ2011,YoshizawaJPSJ2012} 
The latter group pointed out that the Co concentration dependence of the $C_{66}$ softening is ascribable to the presence of a ``structural QCP'', similar to a magnetic QCP, and suggested that the high-$T_c$ in \BFCA \ is related to the structural QCP.\cite{YoshizawaJPSJ2012} 
Since the temperature dependence of $C_{66}$ is quite similar to that of $(T_1T)^{-1}_{\rm AF}$ of AF spin fluctuations, we plot the temperature dependence of $C_{66}$ against that of $(T_1T)^{-1}_{\rm AF}$ with $T$ as an implicit parameter, as shown in Fig.~\ref{C66vsT1T}. 
An apparent proportionality between the two quantities strongly suggests that AF spin fluctuations and structural fluctuations are closely related, indicative of sharing the same origin.

\begin{figure}[tb]
\begin{center}
\includegraphics[width=8cm]{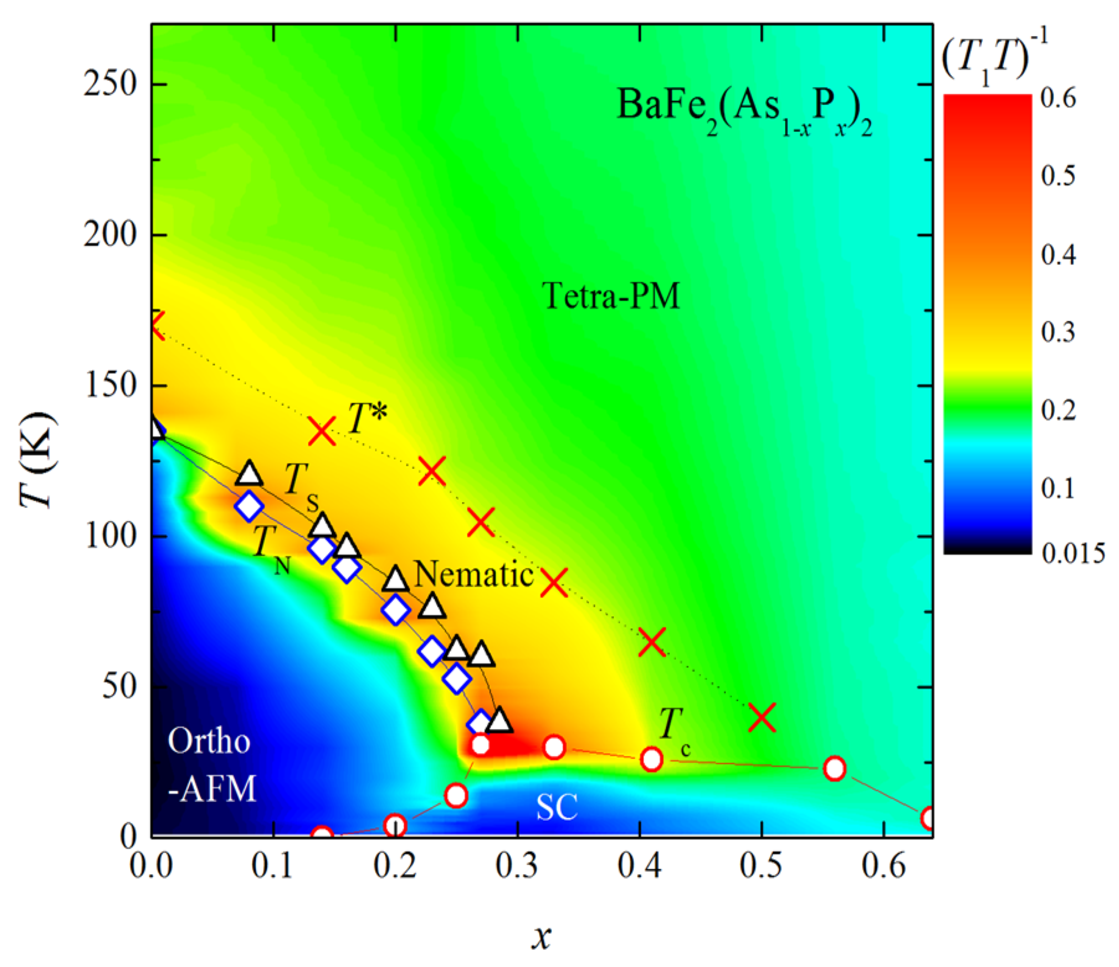}
\end{center}
\caption{(Color online) The nematic transition temperature $T^*$ and $(T_1T)^{-1}$ are compared. The magnitude of $(T_1T)^{-1}$ is shown as a contour plot. $(T_1T)^{-1}$ seems to increase below $T^{*}$, particularly in the low P concentration, suggestive of a close relationship between $T^*$ and AF spin fluctuations.  }
\label{nematicity}
\end{figure}
Quite recently, Kasahara {\it et al.} reported from the torque magnetometry and precise x-ray measurements that the four-fold symmetry is broken in \BFAP\ below $T^*$ that is much higher than $T_S$, and suggested the formation of an electronic nematic state below $T^*$.~\cite{KasaharaNature2012} 
We compare temperature and P concentration dependencies of $T^*$ and $(T_1T)^{-1}$ in Fig.~\ref{nematicity}, where the values of $(T_1T)^{-1}$ are shown as a contour plot.
It seems that $(T_1T)^{-1}$ starts to be enhanced approximately below $T^*$, particularly obvious in the low concentration region. 
The enhancement below $T^*$ can be understood by the fact that the development of the stripe AFM correlations should be determined by breaking the four-fold symmetry in the tetragonal phase, resulting in that the direction of the stripe correlations is fixed and that the correlation length is allowed to be extended more easily. 

The iron-pnictide compounds are thus a unique system, where the spin and orbital degrees of freedom are strongly coupled with each other.
Although we indicated here that the spin-fluctuation theory is successfully applicable to the 122 systems, the interplay between the spin fluctuation and the orbital degrees of freedom remains to be solved in the future.
                          
\section{Conclusion}
We show that the temperature dependencies of the NMR nuclear spin-lattice relaxation rate, the electrical resistivity, and the inelastic neutron scattering data in the paramagnetic phase of iron-pnictide superconductors \BFAP \  and \BFCA \  can be understood quantitatively in the framework of the SCR theory.
A consistent description of these physical properties of \BFAP \  and \BFCA \  in the framework of the SCR theory suggests that an itinerant picture is at work for the ``122'' iron-pnictide superconductors at the low energy scale and AF quantum criticality would be deeply related to the high-$T_c$ superconductivity, as in other unconventional superconductors.
However, a puzzling question in iron-pnictide superconductors is whether AF spin fluctuations and superconductivity are deeply related in ``1111'' systems such as LaFeAs(O$_{1-x}$F$_x$) and Ca(Fe$_{1-x}$Co$_x$)AsF.~\cite{NakaiJPSJ2008,SKitagawaAnisotropy,KobayashiJPSJ2010,OkaPRL2012,TsutsumiPRB2012} 
In addition, the phase diagram in these systems is different from that in the Ba-``122'' systems. Indeed, it was reported recently that superconductivity in LaFeAs(O$_{1-x}$H$_x$) possesses a two-maximum structure and survives until higher hole concentration.~\cite{IimuraNatComm2012} 
It seems that the superconductivity can be observed in the region away from the AF QCP, indicating that the scenarios of superconductivity induced by AF spin fluctuations may not be applied universally. 
Whether a unified picture exists for explaining all experimental results in iron-pnictide superconductors, or whether there exist mechanisms other than magnetism are a future important issue to be clarified. 

\section*{Acknowledgments}
We are grateful to S.~Yonezawa and Y. Maeno for fruitful discussions. 
This work is supported by the Grants-in-Aid for Scientific Research on Innovative Areas ``Heavy Electrons" (No.~20102006) from MEXT, for the GCOE Program ``The Next Generation of Physics, Spun from Universality and Emergence" from MEXT, and for Scientific Research from JSPS. 
Y.N. is supported by KAKENHI (No.~23654120).

\end{document}